# Numerical Predictions of Effective Thermal Conductivities for Three-dimensional Four-directional Braided Composites Using the Lattice Boltzmann Method

Wen-Zhen Fang, Jian-Jun Gou, Hu Zhang, Li Chen, Wen-Quan Tao*
Key Laboratory of Thermo-Fluid Science and Engineering, Ministry of Education, China
School of Energy & Power Engineering, Xi'an Jiaotong University, Shaanxi 710049, PR China
*Corresponding author: wqtao@mail.xjtu.edu.cn

**Abstract**     In this paper, a multiple-relaxation-time lattice Boltzmann model with an off-diagonal collision matrix was adopted to predict the effective thermal conductivities of the anisotropic heterogeneous materials whose components are also anisotropic. The half lattice division scheme was adopted to deal with the internal boundaries to guarantee the heat flux continuity at the interfaces. Accuracy of the model was confirmed by comparisons with benchmark results and existing simulation data. The present method was then adopted to numerically predict the transverse and longitudinal effective thermal conductivities of three-dimensional (3D) four-directional braided composites. Some corresponding experiments based on the Hot Disk method were conducted to measure their transverse and longitudinal effective thermal conductivities. The predicted data fit the experiment data well. Influences of fiber volume fractions and interior braiding angles on the effective thermal conductivities of 3D four-directional braided composites were then studied. The results show that a larger fiber volume fraction leads to a larger effective thermal conductivity along the transverse and longitudinal directions; a larger interior braiding angle brings a larger transverse thermal conductivity but a smaller one along the longitudinal direction. It is also shown that for anisotropic materials the periodic boundary condition is different from the adiabatic boundary condition and for periodic microstructure unit cell the periodic boundary condition should be used.

## 1. Introduction

In recent years, anisotropic materials have been widely used in many engineering applications. Heat and mass diffusions in anisotropic materials have preferable directions, and it is rather difficult to obtain the analytical solutions, especially for the composite materials whose components are anisotropic [1]. The effective thermal conductivity is an important parameter that can quantitatively evaluate the heat transfer capacity in composites. The lattice Boltzmann method (LBM) is an effective approach to solve the Navier-Stokes Equations, and it has been widely used to simulate many kinds of fluid flow problems [2-5]. LBM is a mesoscopic numerical method based on the evolution of particle distribution functions. The behavior of distribution functions is similar to the particle, and its evolution procedure at each grid can be divided into two steps: collision and streaming. Such characteristics make it being able to easily implement complex boundary conditions and multiphase interactions, and can guarantee the conservations without many additional efforts [6]. Attempts have also been made to solve the energy transport equation by LBM, and it has achieved considerable success. Some examples for solving heat transfer or mass diffusion problem by LBM are provided below. Xuan et al. [7] investigated the mass transfer process of volatile organic compounds in porous media based on the LBM. Chen et al. [8] adopted the LBM to calculate the effective diffusivity of the porous gas diffusion layer. Wang et al. [9] proposed a LB algorithm to deal with the fluid-solid conjugate heat transfer problem, which can ensure the heat flux and temperature continuity at the interfaces.

For isotropic heterogeneous materials, much research has been conducted on the predictions of their transport property. In particularly, Wang et al. [10] proposed a LB model to calculate the effective thermal conductivity for granular structure, netlike structure and fibrous structure composite materials. However, this method is restricted to the situation that each component in composites must be isotropic. This is because that the original LB model, the Bathnagar-Gross-Krook model, has only a single relaxation time coefficient. The single-relaxation-time model has insufficient parameters to fully describe the anisotropic heat transfer problem. Several studies have been conducted on the solution of anisotropic heat transfer equation. Zhang et al. [11, 12] proposed a LB model in which the relaxation time coefficients are assumed to be directionally dependent and ensure that the collision is mass-invariant; Ginzburg et al. [13] presented two LB models, the equilibrium-type and the link-type model, to solve the anisotropic heat transfer problems. Although those models can deal with the anisotropic heat transfer situations, they still suffer from the defects in stability and poor application flexibility. Recently, The multiple-relaxation-time (MRT) LB model has been the focus of much research due to its higher stability and accuracy than the single relaxation time model [14, 15], and particularly Yoshida and Nagaoka [16] proposed a developed MRT LB scheme whose collision operator with off-diagonal components enables us to solve the convection and anisotropic heat transfer problem. Although the MRT model for the anisotropic heat transfer problem has been developed, it is only suitable for the homogeneous problem. As for the heterogeneous materials whose components are anisotropic, it will lead to heat flux discontinuity at the interfaces if no additional treatment was adopted. The three-dimensional (3D) four-directional braided composites are typically anisotropic materials, and they are composed of the matrix and braiding yarns. They have been widely applied in aeronautics and astronautics due to their rather high strength and low density [17]. The braiding yarns, one component in 3D four-directional braided composites, are anisotropic with different thermal conductivities along the transverse and longitudinal directions [18,19]. Although several investigations on the effective thermal conductivity of 3D four-directional braided composites had been studied by finite-element method in previous studies [19, 20, 21], the comparisons of the predicted values and the corresponding experimental data were seldom reported. To the authors' knowledge, predictions of the effective thermal conductivities for the anisotropic heterogeneous materials whose components are also anisotropic (for example 3D four-directional braided composites) using the LBM have not been reported in open published literature. The method of the MRT model developed by Yoshida and Nagaoka and the treatment for the internal boundaries should be combined to deal with such heterogeneous materials with anisotropic components. The single-relaxation-time LBM provided by Wang et al. [10] is only suitable for the heterogeneous materials with isotropic components, and it has been used to predict the effective thermal conductivity of the overall isotropic composites with isotropic composites [22] and the directional effective thermal conductivity of the overall anisotropic materials with isotropic components [23].

In the present paper, a multiple-relaxation-time LB model combined with the 'half lattice division scheme' treatment for internal boundaries was adopted to predict the effective thermal conductivity of the anisotropic heterogeneous materials whose components are also anisotropic. With the 'half lattice division scheme' method, the temperature and heat flux can be directly obtained based on the local particle distribution functions without the calculations of finite-difference, and it is important for the continuity of temperature and heat flux at the interfaces (will be discussed at section 2.3). It is worth mentioning that the present method has the potential to deal with the heterogeneous materials with random distributed anisotropic composites, like needled C/SiC composites [24], which currently is still challenging to be solved by finite-element method. In addition, to verify the accuracy and reasonableness of the present method, some simple numerical tests are first performed, and then corresponding experiments are conducted based on the

Hot Disk method to measure the effective thermal conductivity of 3D four-directional braided composites. The influences of fiber volume fractions and interior braiding yarns on the effective thermal conductivity are also discussed in this paper. Estimations of the effective thermal conductivity of the composites are very important for their engineering applications. It provides a guideline for the new materials design. One can even reproduce the structure of the new materials, and predict their effective thermal conductivity based on the method presented in this paper.

## 2. Numerical method

### 2.1 Governing equation

Heat conduction in anisotropic materials has preferable directions due to their directionally dependent thermal conductivity. The governing equations for anisotropic heat conduction in multiphase system, e.g., the matrix and reinforced fibers, without heat sources can be expressed as:

$$\frac{\partial T_m}{\partial t} = \frac{\partial}{\partial x_i}\left((D_{ij})_m \frac{\partial T_m}{\partial x_j}\right) \tag{1}$$

$$\frac{\partial T_f}{\partial t} = \frac{\partial}{\partial x_i}\left((D_{ij})_f \frac{\partial T_f}{\partial x_j}\right) \tag{2}$$

where the subscript $m$ represents the matrix and $f$ represents the reinforced fiber, $T$ denotes the temperature, and $D_{ij}$ is the thermal diffusivity matrix.

The conditions of internal interfaces between different components (phases) must satisfy the continuity of temperature and normal heat flux at the interface, and it can be expressed as [25]:

$$T_m = T_f \tag{3}$$

$$-n_i (\lambda_{ij})_m \frac{\partial T_m}{\partial x_j} = -n_i (\lambda_{ij})_f \frac{\partial T_f}{\partial x_j} \tag{4}$$

where $n_i$ is the unit normal vector at the interfaces, and $\lambda_{ij}$ is the thermal conductivity matrix.

### 2.2 MRT lattice Boltzmann model

The MRT LB model with a collision operator matrix has sufficient parameters to take account of the fully anisotropic heat transfer problems. A new concept, namely moment, is adopted in the MRT model, and it can be obtained from the particle distribution functions. Different from the single-relaxation-time model, the collision step of DnQb MRT model is carried out on a b-dimension moment space, and each moment can be relaxed to the equilibrium state with a different coefficient [15]. While the streaming step of DnQb MRT model is still performed on the velocity space. The velocity space consists of the particle distribution functions, $f_1, f_2, \cdots, f_b$.

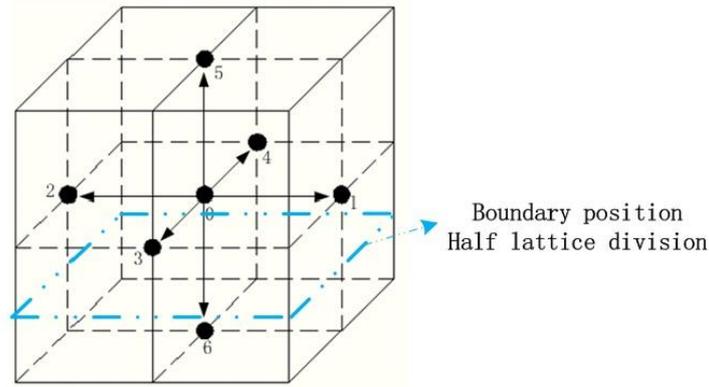

Fig. 1 D3Q7 model scheme

We adopt the three-dimensional seven-speed (D3Q7) MRT model (see Fig. 1) to deal with the anisotropic heat conduction problems. For a pure conduction in the each phase of composite materials, the evolution equation for the MRT LB model that governs the behavior of particle distribution functions can be expressed as [26]:

$$\boldsymbol{f}(\mathbf{x}+\mathbf{e}_i \delta t, t+\delta t) - \boldsymbol{f}(\mathbf{x},t) = -\boldsymbol{\Omega}(\boldsymbol{f} - \boldsymbol{f}^{\text{eq}}) \tag{5}$$

where $\mathbf{x}$ denotes the particle position, $t$ is the real time, $\delta t$ is the time step, $\boldsymbol{f}$ is the particle distribution function vector with seven components, denoted by $\boldsymbol{f} = (f_0, f_1, f_2, f_3, f_4, f_5, f_6)^{\text{T}}$, and for the each component of vector $\boldsymbol{f}^{\text{eq}}$, $f_i^{\text{eq}}$, is the corresponding equivalent particle distribution functions given by

$$f_i^{eq} = \begin{cases} (1-6\gamma)T, & i=0 \\ \gamma T, & i=1,2,\cdots,6 \end{cases} \tag{6}$$

where $\gamma \in (0, 1/6)$ is a constant. $\mathbf{e}_i$ is the discrete velocity, defined as

$$[\mathbf{e}_i] = [\mathbf{e}_0, \mathbf{e}_1, \mathbf{e}_2, \mathbf{e}_3, \mathbf{e}_4, \mathbf{e}_5, \mathbf{e}_6] = \begin{bmatrix} 0 & 1 & -1 & 0 & 0 & 0 & 0 \\ 0 & 0 & 0 & 1 & -1 & 0 & 0 \\ 0 & 0 & 0 & 0 & 0 & 1 & -1 \end{bmatrix} c \tag{7}$$

and $\Omega$ is the collision matrix, defined as

$$\Omega = \mathbf{M}^{-1}\mathbf{S}\mathbf{M} \tag{8}$$

here, $\mathbf{S}$ is a relaxation time matrix, and $\mathbf{M}$ is a matrix that linearly transforms the velocity space into the moment space:

$$\mathbf{M} = [\boldsymbol{\phi}_1, \boldsymbol{\phi}_2, \cdots, \boldsymbol{\phi}_7]^{\mathrm{T}}, \tag{9}$$

$$\boldsymbol{m} = \mathbf{M} \cdot \boldsymbol{f} \tag{10}$$

where $\boldsymbol{\phi}_k$ are the orthogonal basic vectors which are the polynomial functions of the velocity. $\boldsymbol{m}$ is the moment vector with seven components, and the seven moments constitute a b-dimension moment space. The definitions of the matrix $\mathbf{M}$ and the relaxation time matrix $\mathbf{S}$ are defined as [16]:

$$\mathbf{M} = \begin{bmatrix} 1, & 1, & 1, & 1, & 1, & 1, & 1 \\ 0, & 1, & -1, & 0, & 0, & 0, & 0 \\ 0, & 0, & 0, & 1, & -1, & 0, & 0 \\ 0, & 0, & 0, & 0, & 0, & 1, & -1 \\ 6, & -1, & -1, & -1, & -1, & -1, & -1 \\ 0, & 2, & 2, & -1, & -1, & -1, & -1 \\ 0, & 0, & 0, & 1, & 1, & -1, & -1 \end{bmatrix} c \tag{11}$$

$$\mathbf{S}^{-1} = \begin{bmatrix} \tau_0, & 0, & 0, & 0, & 0, & 0, & 0 \\ 0, & \tau_{xx}, & \tau_{xy}, & \tau_{xz}, & 0, & 0, & 0 \\ 0, & \tau_{yx}, & \tau_{yy}, & \tau_{yz}, & 0, & 0, & 0 \\ 0, & \tau_{zx}, & \tau_{zy}, & \tau_{zz}, & 0, & 0, & 0 \\ 0, & 0, & 0, & 0, & \tau_4, & 0, & 0 \\ 0, & 0, & 0, & 0, & 0, & \tau_5, & 0 \\ 0, & 0, & 0, & 0, & 0, & 0, & \tau_6 \end{bmatrix} \tag{12}$$

For isotropic heat conduction problem, $\tau_{xx} = \tau_{yy} = \tau_{zz}$, and $\tau_{ij} = 0$ ($i \neq j$). The off-diagonal components of the relaxation time matrix enable us to take account of the fully anisotropic thermal conduction situation. For each phase in composite materials, the relationships between the relaxation time coefficients and thermal conductivity matrix can be expressed by [16, 27]:

$$\left(\tau_{ij}\right)_m = \frac{1}{2}\delta_{ij} + \frac{\left(D_{ij}\right)_m}{\varepsilon c^2 \delta t}, \quad i, j = 1, 3$$

$$\left(\tau_{ij}\right)_f = \frac{1}{2}\delta_{ij} + \frac{\left(D_{ij}\right)_f}{\varepsilon c^2 \delta t}, \quad i, j = 1, 3$$

(13)

where $\delta_{ij}$ is the Kronecker symbol, and $c$ is the pseudo sound speed, whose value should ensure the value of $\tau_{ii}$ ($i$=1, 2, 3) between 0.5 and 2 [22]. A larger value of $c$ will result in a higher accuracy but a slow rate of convergence [22]. The relaxation time coefficients of $\tau_0$, $\tau_4$, $\tau_5$, $\tau_6$ are generally set to be unity without affecting the numerical results. If the equivalent distribution functions are as defined in Eq. (6), $\varepsilon$ equals $2\gamma$. In this paper, the $\gamma$ is set to be 1/8 [16].

To recover the evolution equation of discrete distribution functions to macroscopic Navies-stokes equations, the fourth-order tensor of discrete velocities, $\sum e_{\alpha i} e_{\alpha j} e_{\alpha k} e_{\alpha l}$, should be isotropic. The energy transport equation has no isotropy requirement of the fourth-order tensor of discrete velocities, making it possible to minimize the numbers of discrete velocities. Therefore, we adopt the D3Q7 model instead of D3Q15 or D3Q19 model to reduce the calculation time without affecting the accuracy. The scheme stated above is second order accuracy with respect to the lattice interval $\delta x$ and first order accuracy with respect to the time step $\delta t$ [16].

**2.3 Internal interfaces and boundary condition treatment**

At the internal interfaces between two different components (phases), the restriction conditions expressed in Eqs. (3-4) must be satisfied to ensure the continuity of both temperature and heat flux at the interfaces. In the LBM, such conditions can be satisfied if we follow the 'half lattice division scheme' (see Fig. 1) first proposed by Wang et al in [8] and then further demonstrated in [25]. The half lattice division scheme means that the straight interface position is placed at the middle of two lattice nodes. Thus, one only needs to identify the local property being phase A or phase B, and the temperature and heat flux can be obtained from the local particle distribution functions without any nearby nodes information. In the present paper, the half lattice division scheme is extended to deal with the internal interfaces to guarantee the continuity of the heat flux for an anisotropic heat transfer problem

The local macroscopic temperature can be obtained by the summation of the discrete distribution functions [22]:

$$T = \sum_i f_i \tag{14}$$

Yoshida and Nagaoka [16] provided the following relationships between the local discrete distribution functions and the first order partial derivatives with respect to temperature as follows:

$$\begin{aligned}
-\frac{1}{\varepsilon \delta x}(f_1 - f_2) &= \tau_{xx}\frac{\partial T}{\partial x} + \tau_{xy}\frac{\partial T}{\partial y} + \tau_{xz}\frac{\partial T}{\partial z} \\
-\frac{1}{\varepsilon \delta x}(f_3 - f_4) &= \tau_{yx}\frac{\partial T}{\partial x} + \tau_{yy}\frac{\partial T}{\partial y} + \tau_{yz}\frac{\partial T}{\partial z} \\
-\frac{1}{\varepsilon \delta x}(f_5 - f_6) &= \tau_{zx}\frac{\partial T}{\partial x} + \tau_{zy}\frac{\partial T}{\partial y} + \tau_{zz}\frac{\partial T}{\partial z}
\end{aligned} \tag{15}$$

One can obtain the first order partial derivatives with respect to temperature, $\partial T/\partial x$, $\partial T/\partial y$, $\partial T/\partial z$, by solving the above ternary linear equations. Then, the heat fluxes along the specified directions can be calculated by

$$\begin{aligned}
q_x &= \rho c_p \left( D_{xx}\frac{\partial T}{\partial x} + D_{xy}\frac{\partial T}{\partial y} + D_{xz}\frac{\partial T}{\partial z} \right) \\
q_y &= \rho c_p \left( D_{yx}\frac{\partial T}{\partial x} + D_{yy}\frac{\partial T}{\partial y} + D_{yz}\frac{\partial T}{\partial z} \right) \\
q_z &= \rho c_p \left( D_{zx}\frac{\partial T}{\partial x} + D_{zy}\frac{\partial T}{\partial y} + D_{zz}\frac{\partial T}{\partial z} \right)
\end{aligned} \tag{16}$$

where $\rho c_p$ is the volume specific capacity.

In the conventional conjugate heat transfer problem, the general governing equation adopted for discretization in its vector form is as follows [28, 29]:

$$\frac{\partial(\rho\phi)}{\partial t} + div(\rho\vec{U}\phi) = div(\Gamma_\phi grad\phi) + S_\phi \tag{17}$$

where $\phi$ is a general dependent variable to be solved, and $\Gamma_\phi$ the related nominal diffusion coefficient, which is defined by the following equation:

$$\Gamma_\phi = \frac{\lambda}{c_p} \tag{18}$$

Then as first pointed out by Chen and Han in [30] and later further demonstrated in [28], for a conjugated heat transfer problem if the interface diffusion coefficient is determined by the harmonic mean, the specific heat capacity ($c_p$) in the solid region should take the value of fluid in order to guarantee the continuity of flux at the interface.

The LBM model adopted in this paper has its energy equation of the following form:

$$\frac{\partial T}{\partial t} + div(\vec{U}T) = div(D_T gradT) + S_T \tag{19}$$

Thus, according to above discussions, we should assume:

$$(\rho c_p)_m = (\rho c_p)_f = 1 \tag{20}$$

for different phases in composites to ensure the heat flux continuity [31]. Such treatment restricts the application of the present model only for the steady case. When the heat transfer reaches the steady state, it does not influence the temperature field. Based on this assumption (shown in Eq. (20)), the incoming distribution functions at the interfaces can be obtained through the streaming process without any additional treatment:

$$\begin{aligned} f_{\bar{\alpha}}(\mathbf{x}_f, t+\delta t) &= \hat{f}_{\bar{\alpha}}(\mathbf{x}_m, t) \\ f_\alpha(\mathbf{x}_m, t+\delta t) &= \hat{f}_\alpha(\mathbf{x}_f, t) \end{aligned} \tag{21}$$

Note that if the heat flux continuity at the internal interfaces needs to be satisfied in a transient analysis, the incoming distribution functions at the interfaces need to be modified to ensure the heat flux continuity at the transient problem, and therefore the streaming process should be modified at the interfaces. One can refer to references [25, 27] for more details. In the present paper, only the steady state heat transfer problem was studied because it is sufficient in estimating the effective thermal conductivity of the materials.

Once the temperature field is solved, the effective thermal conductivity along specified directions can be calculated by

$$\lambda_{x,eff} = \frac{L_x \cdot \int q_x dA_x}{\Delta T \cdot A_x}$$
$$\lambda_{y,eff} = \frac{L_y \cdot \int q_y dA_y}{\Delta T \cdot A_y} \quad (22)$$
$$\lambda_{z,eff} = \frac{L_z \cdot \int q_z dA_z}{\Delta T \cdot A_z}$$

where $L_x$, $L_y$, $L_z$ are the thicknesses of materials along x, y, z directions, respectively; $q_x$, $q_y$, $q_z$ are steady heat fluxes along the x, y, z directions, respectively; and $\Delta T$ is the temperature difference between the two opposite surfaces.

Boundary conditions for the LB simulations are as follows. For the unit cube cell of materials, two opposite boundary surfaces are set to be isothermal but at different temperatures (Dirichlet condition). Other surfaces are set to be adiabatic (Neumann condition) or periodic according to the actual situation. For the position placed at the middle of two lattice node (see Fig. 1), the following treatments for Dirichlet and Neumann conditions have the second order accuracy [16, 32].

Dirichlet condition:

$$f_\alpha(\mathbf{x}, t+\delta t) = -\hat{f}_{\bar{\alpha}}(\mathbf{x},t) + \varepsilon \phi_d \quad (23)$$

Neumann condition:

$$f_\alpha(\mathbf{x}, t+\delta t) = \hat{f}_{\bar{\alpha}}(\mathbf{x},t) + (\delta t / \delta x)\phi_n \quad (24)$$

While the periodic condition is set as follows:

$$f_\alpha(\mathbf{x}+\mathbf{L}, t+\delta t) = \hat{f}_\alpha(\mathbf{x}, t) \tag{25}$$

here $\hat{f}$ denotes the post-collision discrete distribution function, the index $\bar{\alpha}$ indicates the directions opposite to $\alpha$, $\phi_d$ is the given temperature and $\phi_n$ is the given specified flux at the boundary.

## 3. Validation test

In this section, some benchmark problems are simulated to validate the accuracy of the method in Section 2.

### 3.1 Infinite anisotropic thin slabs

A series mode of two anisotropic thin slabs is considered, and their effective thermal conductivity can be obtained analytically if the geometry size along the x direction is infinite long (see Fig. 2). In Fig. 2, the oblique lines represent the principle axis of heat conduction, and $\beta$ is the oblique angle. The thermal conductivities along the two principle axes of heat conduction are denoted as $\lambda_\eta$ and $\lambda_\zeta$, respectively. To obtain the effective thermal conductivity of the composites composed of two anisotropic thin slabs in series, the effective thermal conductivity of single slab along y direction, $\lambda_y$, should be first determined. The thermal conductivity matrix of the anisotropic slab in $\eta$–$\zeta$ coordinate is diagonal, but it should be converted into x-y coordinate. The off-diagonal components of the thermal conductivity matrix in x-y coordinate originate from the rotation of the principle axes

$$\begin{bmatrix} \lambda_{xx} & \lambda_{xy} \\ \lambda_{yx} & \lambda_{yy} \end{bmatrix} = \begin{bmatrix} \cos\beta & \sin\beta \\ -\sin\beta & \cos\beta \end{bmatrix} \begin{bmatrix} \lambda_\eta & 0 \\ 0 & \lambda_\zeta \end{bmatrix} \begin{bmatrix} \cos\beta & -\sin\beta \\ \sin\beta & \cos\beta \end{bmatrix} \tag{26}$$

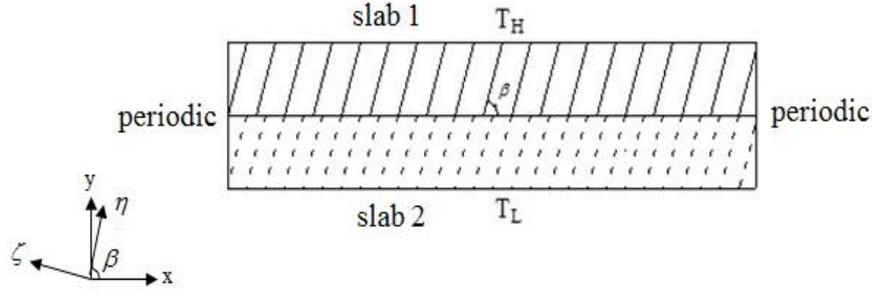

Fig. 2 Series mode of two infinite anisotropic thin slabs

And the heat flux can then be expressed as

$$\begin{bmatrix} q_x \\ q_y \end{bmatrix} = - \begin{bmatrix} \lambda_{xx} & \lambda_{xy} \\ \lambda_{yx} & \lambda_{yy} \end{bmatrix} \begin{bmatrix} \partial T/\partial x \\ \partial T/\partial y \end{bmatrix} \qquad (27)$$

If the slab is infinite long along the x direction, $\partial T/\partial x$ equals zero and therefore the effective thermal conductivity of the single slab along y direction, $\lambda_y$, is equal to $\lambda_{yy}$. If we assume the two slabs have the same thickness, the effective thermal conductivity of the series slabs is equal to $2\lambda_{y1}\lambda_{y2}/(\lambda_{y1}+\lambda_{y2})$ ( here, index 1 and 2 denote two different slabs).

Table I Comparisons of the predicted results and the analytical results

| | $\beta=75°$ | | | $\beta=15°$ | | |
|---|---|---|---|---|---|---|
| | Analytical results | Predicted results | Relative deviations | Analytical results | Predicted results | Relative deviations |
| $\lambda_{\eta 1}:\lambda_{\eta 2}$ | (W/(m·K)) | (W/(m·K)) | (%) | (W/(m·K)) | (W/(m·K)) | (%) |
| 1:1 | 1.9330 | 1.9330 | 0 | 1.0670 | 1.0670 | 0 |
| 1:10 | 3.5145 | 3.5145 | 0 | 1.9400 | 1.9401 | 0.005 |
| 1:20 | 3.6819 | 3.6820 | 0.003 | 2.0324 | 2.0324 | 0 |
| 1:40 | 3.7717 | 3.7717 | 0 | 2.0820 | 2.0821 | 0.005 |
| 1:100 | 3.8277 | 3.8278 | 0.003 | 2.1129 | 2.1129 | 0 |
| 1:1000 | 3.8621 | 3.8622 | 0.003 | 2.1319 | 2.1320 | 0.005 |

The effective thermal conductivities predicted by the present method and the corresponding analytical results are shown in the Table I. We keep the ratio $\lambda_\eta = 2\lambda_\zeta$ for both slabs, and set $\lambda_{\eta 1}=2$ W/(m·K) while change the value of $\lambda_{\eta 2}$ from 2 to 2000 W/(m·K), that is, the ratio of $\lambda_{\eta 2}/\lambda_{\eta 1}$ varying from one to one thousand. In our simulations, the upper and lower boundaries are set to be isothermal but at different constant temperature. The periodic conditions are imposed on the side boundary to satisfy the infinity assumption. The size of grid space is 0.01 at a 200×200 grid, and the value of pseudo sound speed $c$ is maintained to be 400000. The maximum relative deviation is 0.003% for $\beta=75°$ and 0.005% for $\beta=15°$, which confirms the high accuracy of the present approach.

**3.2 Composites reinforced with anisotropic short fibers**

Reinforced fibers are commonly dispersed in the solid matrix to satisfy the demand for their applications [33]. We consider that the reinforced short orthotropic fibers with transverse isotropy, $\lambda_{xx} \neq \lambda_{yy} = \lambda_{zz}$, are longitudinally aligned in the matrix, as shown in Fig. 3.

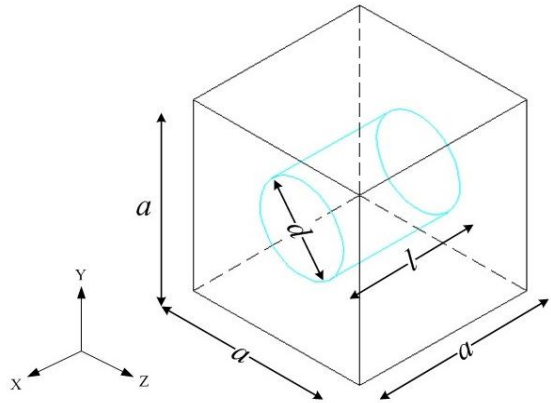

Fig. 3 Cubic-cell geometry reinforced with short fiber

The fiber volume fraction, $v_f$, and the fiber aspect ratio, $\gamma$, are defined as

$$v_f = \frac{\pi d^2 l}{4a^3} \tag{28}$$

$$\gamma = \frac{l}{d} \tag{29}$$

where $a$ is the side length of the cubic; $d$ is the fiber diameter; $l$ is the fiber length. For a given fixed fiber volume fraction, $v_f$, there exists a minimum fiber aspect ratio, $\gamma_{\min}$, and a maximum value, $\gamma_{\max}$ [33].

$$\gamma_{\min} = \frac{4v_f}{\pi}, \quad \gamma_{\max} = \sqrt{\frac{\pi}{4v_f}} \tag{30}$$

To study the influences of the fiber aspect ratio on the effective thermal conductivity of composites, we keep the fiber volume fraction at a fixed value while change the value of fiber aspect ratio. In Ref. [33], the effective thermal conductivities of such composites are numerically estimated by the finite element method. We set the same parameter values as those in Ref. [33]. The matrix material is set to be isotropic and its thermal conductivity is assigned to be unity. The reinforced short fiber is assumed to be transversely isotropic, $\lambda_{yy}=\lambda_{zz}$. The degree of the fiber anisotropy is defined as $\mu = \lambda_{xx}/\lambda_{yy}$. $\lambda_L^e$ and $\lambda_T^e$ are the longitudinal and the transverse effective thermal conductivity of the composites, respectively. Comparisons between the predicted thermal conductivities by the present method and that in Ref. [33] are shown in Fig. 4. In Fig. 4(a), $\lambda_{xx}=$ 1000 W/(m·K) and $\mu=10$ for fiber; while in Fig. 4(b), $\lambda_{xx}=100$ W/(m·K) and $\mu=0.1$. Note that the step-wise approximation is adopted to deal with the curved boundaries. The accuracy of the step-wise approximation depends on the grid resolution. For all the cases, the values of the effective thermal conductivity obtained based on an 80×80×80 grid system are not much different with those obtained based on the 60×60×60 one, and their deviations are within 1.2%. Therefore, the grid system 60×60×60 is enough for this problem. All the results shown in Fig. 4 are obtained based on the 60×60×60 grid. The size of grid space is 0.01 and the value of pseudo sound speed $c$ is maintained as 100000. As shown in Fig. 4, the predicted results by the present method are in good agreement with the existing simulation data, and the maximum deviation is 2.2%, which again confirms the accuracy of the present approach.

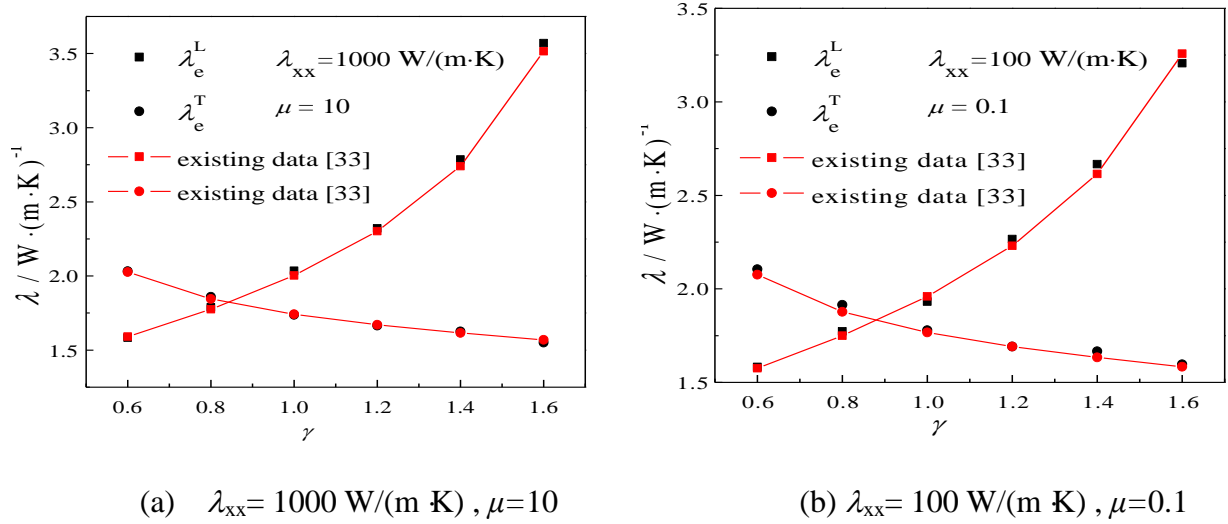

(a) $\lambda_{xx}$= 1000 W/(m·K), $\mu$=10  (b) $\lambda_{xx}$= 100 W/(m·K), $\mu$=0.1

Fig. 4 Comparisons of effective thermal conductivity between the predicted results and the existing simulation data [33]

### 3.3 Three-dimensional dual-component composites

In this section, we only consider one basic structure, series mode of 3D dual-component composites (see Fig. 5).

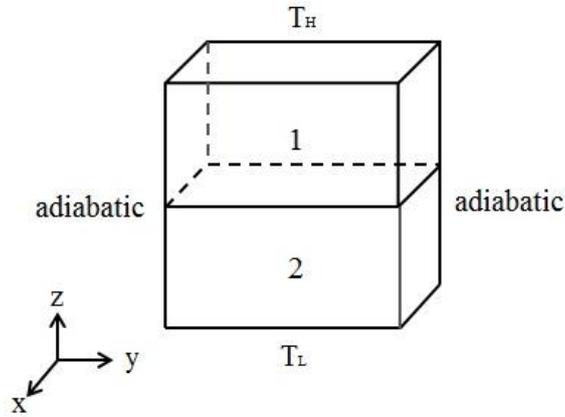

Fig. 5 Series mode of 3D dual-component composites

The thermal conductivity matrices of the each component are given as:

$$\lambda_{ij}^1 = \begin{bmatrix} 4 & 1 & 2 \\ 1 & 8 & 1 \\ 2 & 1 & 10 \end{bmatrix}, \quad \lambda_{ij}^2 = \begin{bmatrix} 10 & 1 & 4 \\ 1 & 8 & 1 \\ 4 & 1 & 12 \end{bmatrix} \tag{31}$$

In our simulations, the boundary conditions are imposed as shown in Fig. 5. Ansys Fluent 14.0 was adopted to obtain the effective thermal conductivity of the composite material along the z direction, and it equals 10.34 W/(m·K). The effective thermal conductivity along the z direction predicted by the present method equals 10.40 W/(m·K) with a grid of 30×30×30. The size of grid space is 0.01 and the value of pseudo sound speed $c$ for LBM model is maintained to be 8000. The deviation is 0.6%. The temperature distribution contours of the right hand side surface of the cube obtained by Ansys Fluent and the present method are compared in Fig. 6, which shows good agreement.

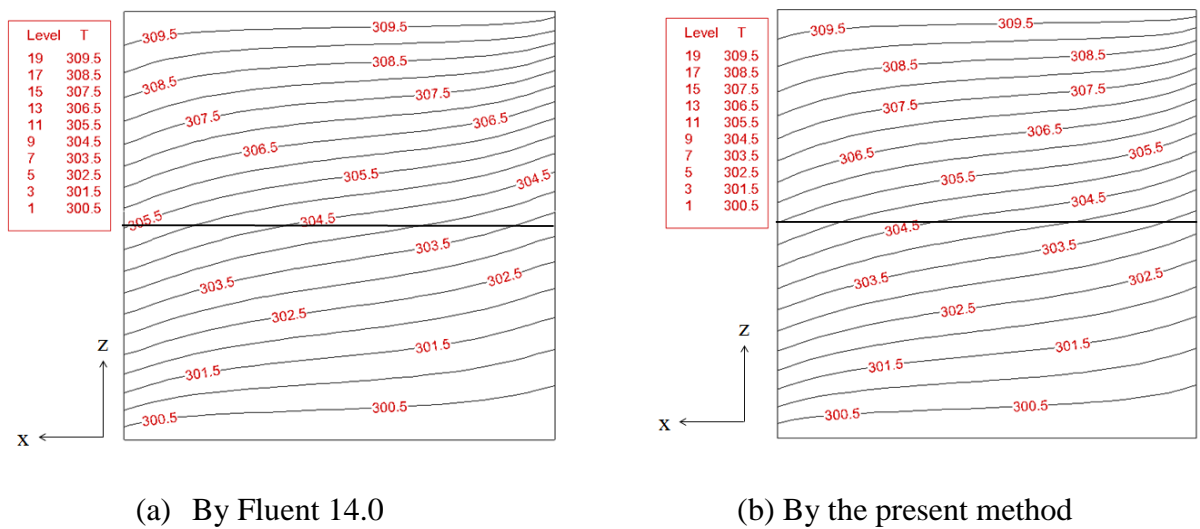

(a) By Fluent 14.0  (b) By the present method

Fig. 6 Temperature distribution contour of the right hand side surface of the cube

It is worth noting that at the interface, the temperature contours obtained either by Fluent or by the present method are continuous. The local normal heat flux can be obtained by Eq. (16), and it is found that normal heat flux at the interface is also continuous with a deviation less than 0.5%. The continuity of temperature and heat flux at the interface verifies the accuracy of the present 'half lattice division scheme' treatment extended for the anisotropic heat conduction problem.

## 4. Application for 3D four-directional braided composites

### 4.1 structure of 3D four-directional braided composites

For 3D four-directional braided composites, the inside braiding yarns are regularly woven by machines and their structure is periodic. A representative unit cell thus can be built to describe the entire composites according to the movement of braiding yarns during the braiding process, and the effective thermal conductivity of braided composites can be obtained based on such representative unit cell [18, 19, 34]. In the present paper, the unit cell developed in [17] is adopted which contains 12 long straight yarns and 8 short yarns. The coordinates and orientation angles of each yarn axis are the same as that in [17, 19]. A schematic of the unit cell and its components is shown in Fig. 7(a). There are two geometric scale levels: first, thousands of uniaxial fibers and the matrix constitute a braiding yarn; second, lots of braiding yarns within the matrix form the braided composites. In Fig. 7, $\gamma$ is the orientation angle of each yarn with z axis, namely interior braiding angle, $h$ denotes the braiding pitch length, and $a$ is the side length of the unit cell. A reconstructed unit cell studied in the present study is shown in Fig. 7 (b). Actually, the braiding yarns in composites contact tightly with each other, and each yarn is subjected to the compressive force by its adjacent yarns. As a result, the cross section of the yarns will be distorted and no longer a circle. In previous studies, different cross sections were studied such as ellipse [18], hexagon [35] and octagon [36]. In the present work, the cross section is assumed to be ellipse. To satisfy the condition that elliptical-section braiding yarns contact tightly with each other, the sizes of the unit cell and related geometry parameters of braiding yarns must obey the following relations [17]:

$$d = a/8 \tag{32}$$

$$c = d\sqrt{3}\cos\gamma \tag{33}$$

$$A = \pi dc = \frac{n_f \times \pi d_f^2 / 4}{\phi_{fy}} \tag{34}$$

$$\phi_y = \frac{(\pi cd) \times h/\cos\gamma \times 8}{a \times a \times h} = \frac{\pi\sqrt{3}}{8} = 0.68 \tag{35}$$

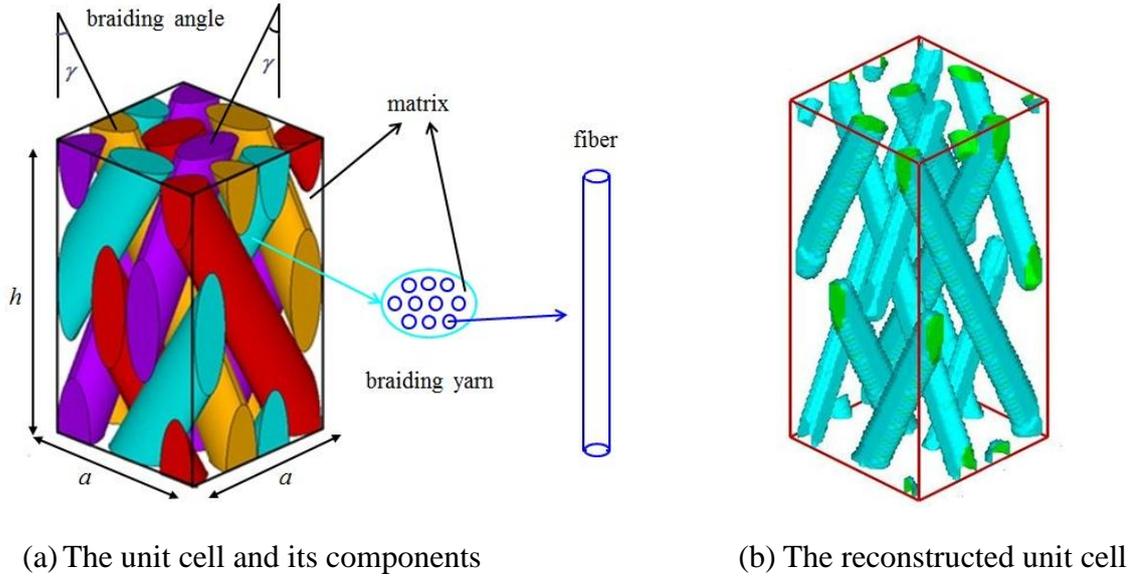

(a) The unit cell and its components          (b) The reconstructed unit cell

Fig. 7 Diagram of the unit cell for 3D four-directional braided composites

where $c$ and $d$ are the lengths of semi-major axis and semi-minor axis of the ellipse cross section, respectively; $A$ is the area of the cross section; $n_f$ is the number of fibers in one braiding yarn; $d_f$ is the diameter of fibers; $\phi_{fy} = \phi_f / \phi_y$ is the fiber volume fraction of the braiding yarns, $\phi_f$ is the fiber volume fraction of the unit cell, and $\phi_y$ is the yarn volume fraction of the unit cell. With values of $\phi_f$, $n_f$, $d_f$ and $\gamma$ given, one can determine the geometry parameter values of $c$, $d$, $a$, $h$ according to the Eqs.(32)-(35). In the present paper, we set $n_f = 9000$, $d_f = 6.9\mu m$, and let $\gamma$ vary from 20 to 45 degrees, $\phi_f$ equals 0.4, 0.5, 0.58. Then the related geometry parameters of a series of unit cells can be determined.

**4. 2 Materials properties and boundary conditions**

To determine the effective thermal conductivities of braided composites, one first has to gain the thermal conductivities of their each component, matrix and braiding yarns. Braiding yarn, one component of braided composites, is composed of the matrix and thousands of uniaxial fibers. In the present paper, the matrix is epoxy resin and the reinforced fiber is T300 carbon. The thermal conductivity of isotropic resin can be experimentally measured by Hot Disk TPS2500s (discussed in

Section 4.3), and equals 0.178 W/(m·K). The T300 carbon fiber is assumed to be transversely isotropic, whose transverse and longitudinal thermal conductivities are 0.675 W/(m·K) and 7.81 W/(m·K), respectively [37]. According to the determined thermal conductivities of the matrix resin and T300 carbon fibers, the longitudinal and transverse thermal conductivities of braiding yarns can be obtained by [21]:

$$\lambda_y^L = \lambda_f^L \phi_{fy} + \lambda_m (1 - \phi_{fy}) \tag{36}$$

$$\lambda_y^T = \lambda_m + \frac{(1 - \phi_{fy})}{1/(\lambda_f^T - \lambda_m) + (1 - \phi_{fy})/(2\lambda_m)} \tag{37}$$

where $\lambda_f^L$ and $\lambda_f^T$ are the longitudinal and transverse effective thermal conductivity of fibers, respectively; $\lambda_m$ is the thermal conductivity of the matrix.

For 3D four-directional braided composites, two opposite surfaces along the measured direction are isothermal but at different given temperatures. The other boundary surfaces are periodic boundary condition [19].

**4.3 Experimental measurement**

In the present paper, a Hot Disk thermal constant analyzer (TPS 2500s) [38-40] based on the transient plane source method was adopted to determine the effective thermal conductivities of materials, including the isotropic resin and the anisotropic braided composites. For 3D four-directional braided composites, the specimens should be cut right so that the probe can be placed perpendicular to the braiding direction. The Hot Disk method can determine the transverse and longitudinal effective thermal conductivities of the measured materials simultaneously.

The measurement process is as follows. The probe is clamped between two identical specimen halves, and then a heat pulse is supplied to the probe during a certain time to generate a dynamic temperature field. The temperature increase of the probe surface is recorded as a function of time.

The temperature response within the specimen is predominantly related to the thermal diffusivity and thermal conductivity of the measured material. By dealing with the recorded temperature curve, one can simultaneously obtain the thermal conductivity and thermal diffusivity of the measured material. For isotropic materials, the temperature increase of the probe surface can be expressed as:

$$\Delta T_s(\tau) = \frac{P_0}{\pi^{3/2} r \lambda} D(\tau) \tag{38}$$

where $P_0$ is the input power, $r$ is the radius of the probe, $\lambda$ is the thermal conductivity of the specimen material, $D(\tau)$ is the dimensionless specific time function, and $\tau$ is the dimensionless time, defined as $\tau = \sqrt{at}/r$, $a$ is the thermal diffusivity of the specimen, and $t$ is the measurement time. The thermal diffusivity can be obtained by a least-squares procedure to obtain a best linear relationship between the $\Delta T_s$ and $D(\tau)$. Finally, the thermal conductivity can be obtained by the slope of this line (Eq. (38)).

As for anisotropic materials, the temperature increase of the probe can be expressed as [37]

$$\Delta T_s(\tau_T) = \frac{P_0}{\pi^{3/2} r (\lambda_L \lambda_T)^{-1}} D(\tau_T) \tag{39}$$

where $\lambda_L$ and $\lambda_T$ are the longitudinal and transverse effective thermal conductivities of the specimen, respectively. Similarly, we can first obtain the thermal diffusivity, $a_T$, along the transverse direction. With a given volume specific capacity $C$, then

$$\lambda_T = C \cdot a_T \tag{40}$$

The longitudinal thermal conductivity of the specimen, $\lambda_L$, can then be obtained through the slope of the line corresponding to Eq. (39).

In this paper, Kapton 5465, a Hot Disk probe, with a radius 3.189 mm was adopted; the output power was set as 0.025 W; and the measurement time was set to be 20 s.

## 5. Results and discussion

The MRT LBM combined with the 'half lattice division scheme' method developed in Section 2 is adopted to determine the longitudinal and transverse effective thermal conductivity of 3D four-directional braided composites. The unit cell reconstructed in Fig.7 is studied. Taking the case with fiber volume fraction of 0.4 and the interior braiding angle of 25° as an example, the geometry parameters of unit cell are as follows. The side length of the unit cell, $a$, is 2.724 mm; the braiding pitch length, $h$, is 5.843 mm; the semi-major axis of ellipse, $c$, is 0.535 mm; the semi-minor axis of ellipse, $d$, is 0.341 mm. The grid number of the unit cell is 56×56×120. In the simulation, the braiding yarns are treated to be uniform with the resin and thousands of uniaxial fibers. The measured thermal conductivity of the resin is 0.178 W/(m·K). The longitudinal and transverse thermal conductivities of the braiding yarns obtained by Eq. (36) and Eq. (37) are listed in Table II.

Table II Effective thermal conductivities of braiding yarns

| $\phi_f$ (%) | $\lambda_y^L$ (W/(m·K)) | $\lambda_y^T$ (W/(m·K)) |
|---|---|---|
| 0.4 | 4.667 | 0.363 |
| 0.5 | 5.790 | 0.445 |
| 0.58 | 6.688 | 0.530 |

The braiding yarns in composites are not parallel to the z axis (see Fig. 5), which implies that the principle axes of heat conduction differ from the Cartesian coordinate directions. To calculate the effective thermal conductivity of the braided composites along x, y, z directions, the thermal conductivity matrix of anisotropic yarns should be rotated into Cartesian system. The off-diagonal components of the thermal conductivity matrix in Cartesian system are the results of the rotation. After reconstructing the unit cell of the composites structure and determining the thermal conductivity matrix of each component, we can then apply the MRT LBM to calculate the effective thermal conductivities of anisotropic heterogeneous materials along a specified direction. Note that the half lattice division scheme was adopted to deal with the internal boundaries. Such treatment can guarantee the heat flux continuity at the interfaces. The heat flux along the z direction is shown in Fig. 8, which verifies the energy conservation of this method.

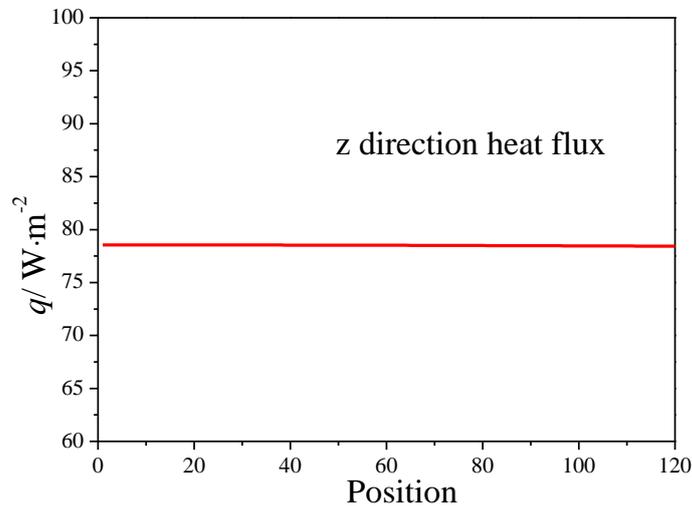

Fig. 8 Heat flux along the z direction

**5. 1 Comparisons with the experimental data**

The Hot Disk thermal constant analyzer presented above was adopted to experimentally measure the effective thermal conductivity to confirm the accuracy and reasonableness of the present method. The experimental data and the predicted results are shown in Table III. In the table $\lambda_T^e$ and $\lambda_L^e$ are the numerically predicted effective thermal conductivities of 3D four-directional braided composites along the transverse and longitudinal directions, respectively, and $\lambda_T$ and $\lambda_L$ are the measured transverse and longitudinal thermal conductivities. The predicted results and the experimental data show good agreements and the deviations are within ±10%. The accuracy of the present numerical method is further confirmed by the experimental data, which thus can be adopted to predict the effective thermal conductivities of the braided composites. In 3D four-directional braided composites, the braiding yarns are regularly woven by machines and therefore their structure is periodic. During the manufacture of the 3D four-directional braided composites, two very important structural parameters, the interior braiding angle and fiber volume fraction of the composites can be precisely controlled. Therefore, the microstructures of the four-directional braided composites are quite regular, which can thus be well reconstructed. Further, due to the periodic structures, a representative unit cell can be extracted to describe the entire composites according to the movement of braiding yarns and is very similar to the real structure of composites. This is exactly the reason that good agreement is obtained between the simulation results based on a representative unit cell and the experimental results based on the entire domain.

Table III Comparisons of the experimental data and predicted results

| $\gamma$ | $\phi_f$ | $\lambda_T^e$ W/(m·K) | $\lambda_T$(Exp) W/(m·K) | deviation % | $\lambda_L^e$ W/(m·K) | $\lambda_L$(Exp) W/(m·K) | deviation % |
|---|---|---|---|---|---|---|---|
| 25 | 0.5  | 0.639 | 0.709 | -9.87 | 3.085 | 3.41 | -9.53 |
| 25 | 0.58 | 0.727 | 0.75  | -3.07 | 3.444 | 3.52 | -2.16 |
| 40 | 0.5  | 1.015 | 1.02  | -0.4  | 2.258 | 2.50 | -9.68 |
| 40 | 0.58 | 1.160 | 1.056 | 9.89  | 2.583 | 2.63 | -1.78 |

In the present paper, all the simulations are based on the assumption that different components in the composites are contacted well tightly with the negligible interface thermal resistance. This assumption is also widely adopted in the literature to study effective thermal conductivity of composite materials. To consider the internal thermal resistance, the morphology of the contacted surface is needed to be specifically described at the micro size level. While the domain size for predicting the effective thermal conductivity is at the millimeter level. Such domain size cannot precisely describe the morphology of the contacted surface. The macroscopic model to consider the thermal resistance using the LBM needs further development. However, the negligible thermal resistance assumption for 3D four-directional braided composites seems reasonable because the numerical results show good agreement with the experimental results. With such an assumption, the present paper focuses on the study of the effects of component contents and component transport properties on the effective thermal conductivity. Currently considering the thermal resistance in numerical simulations is still challenging. It should be mentioned that Wang et al [41] embed the thermal contact resistance into the granular microstructure and Han et al. [42] proposed a partial bounce-back algorithm method in the thermal LBM to account for the thermal contact resistance, but it cannot be used directly in the 3D four-directional braided composites.

**5.2 Influence factor of effective thermal conductivities of braided composites**

The influences of fiber volume fractions and interior braiding angles on the effective thermal conductivities are studied with the fiber volume fractions as 0.4, 0.5, 0.58, and the braiding angles varying from 20 to 45 degrees. The predicted transverse and longitudinal thermal conductivities are shown in Fig. 9 (a) and Fig. 9 (b), respectively. It can be observed that the transverse thermal conductivity increases with the interior braiding angle, while the longitudinal thermal conductivity decreases with the increase in the interior braiding angle. This can be explained as follows: a larger interior braiding angle will lead to a larger proportion of yarns projected into the transverse direction, and a smaller proportion along the longitudinal direction. This means that the heat transfer capability will be strengthened along the transverse direction but reduced along the longitudinal direction due to the bigger effective thermal conductivity of braiding yarns than that in the matrix. It can also be found from the figures that both the transverse and longitudinal effective thermal conductivity increase with the fiber volume fractions within the range of the interior braiding angle studied. This is because the thermal conductivity of fibers is bigger than that in the matrix.

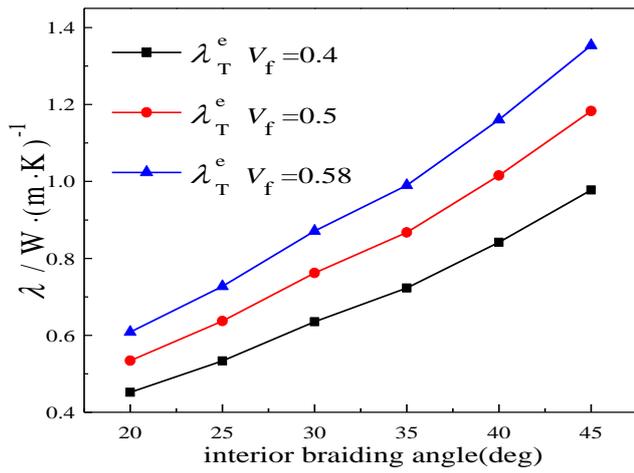
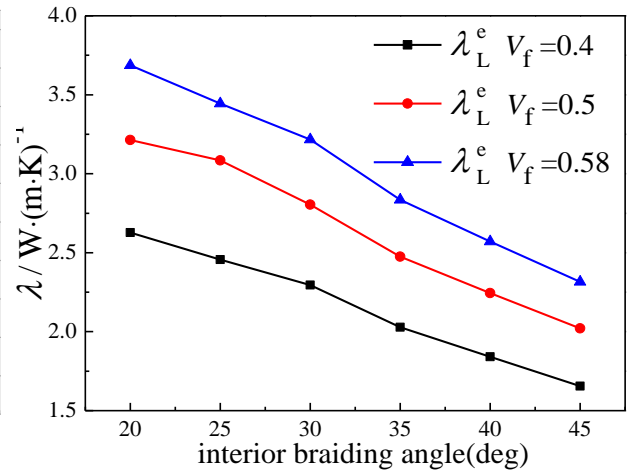

(a)　Transverse thermal conductivities　　　　(b) Longitudinal thermal conductivities

Fig. 9 Effective thermal conductivities versus the interior braiding angle

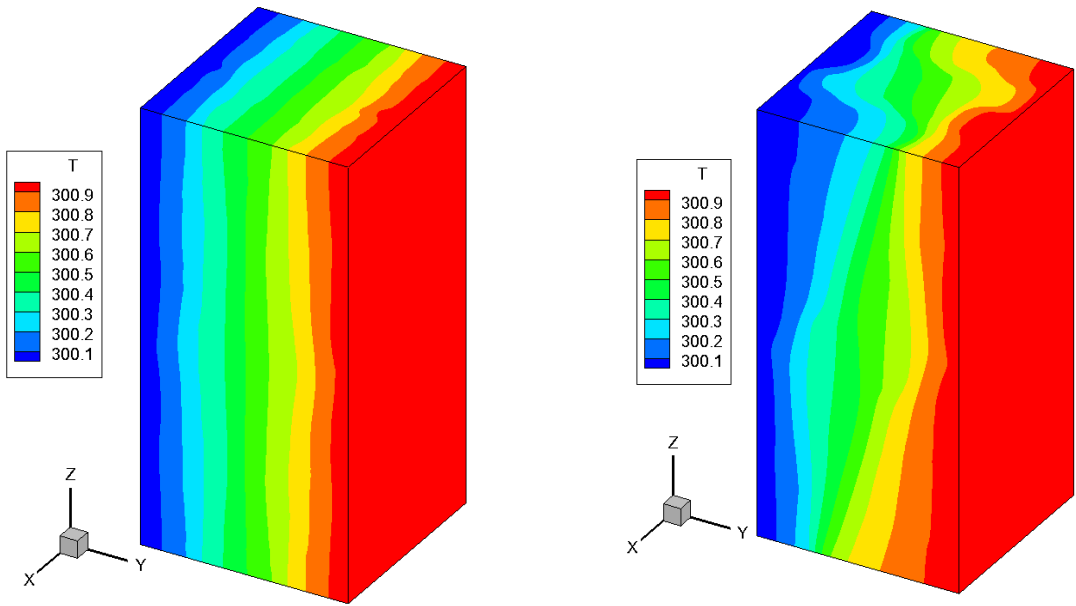

(a)The periodic boundary condition　　　　(b) The adiabatic boundary conditon

Fig. 10 Temperature distribution contours

It is worth mentioning that the above results were obtained based on the periodic boundary condition rather than adiabatic one. For isotropic homogeneous or heterogeneous materials, the imposed boundary condition being adiabatic or periodic will not influence the calculations of effective thermal conductivity. However, as for anisotropic homogeneous or heterogeneous materials, the imposed boundary condition will greatly influence the effective thermal conductivity.

Taking the case $\gamma=25°$ as an example, the transverse and longitudinal thermal conductivities are 0.637 W/(m K) and 3.085 W/(m K) respectively for periodic boundary conditions, while 0.566 W/(m K) and 2.486 W/(m K) for adiabatic boundary conditions. The temperature distribution contours under the two different boundary conditions are shown in Fig. 10. It can be seen that they are much different. The periodic boundary conditions result in less tortuous temperature distribution contours. This is because the periodicity means that the geometry size of measured materials along the imposed direction is infinite, and therefore the influence of the boundary on the temperature field will be reduced. For the anisotropic homogeneous materials, the periodic boundary condition will result in a zero temperature gradient along the imposed direction while the adiabatic boundary condition will not. As for the anisotropic heterogeneous materials, such as the 3D four-directional braided composites, the temperature distribution contours obtained by periodic boundary conditions will be much less tortuous than that of adiabatic one. According to Eq. (16), the temperature gradients of the $x$ or $y$ direction will have an influence on the heat flux along the $z$ direction and therefore result in different effective thermal conductivities along $z$ direction calculated by the boundary condition of two specified wall temperatures. Thus, for periodic microstructure unit cell, we should impose periodic boundary conditions rather than adiabatic boundary conditions.

## 6. Conclusion

In this paper, a multi-relaxation-time LB model combined with the 'half lattice division scheme' method was adopted to predict the effective thermal conductivities of the anisotropic heterogeneous materials whose components are also anisotropic. By benchmark validations and comparisons with the existing simulation data, the accuracy of the present method has been confirmed. This method is then applied to predict the transverse and longitudinal effective thermal conductivity of 3D four-directional braided composites. To check the accuracy of the present method, some corresponding experiments were conducted to measure the effective thermal conductivity of 3D four-directional braided composites. The predicted results agree well with the experimental data. For 3D four-directional braided composites, it is found that both the longitudinal and transverse thermal conductivity increase with the fiber volume fraction; the transverse thermal conductivity of the braided composites increases with the interior braiding angle while the longitudinal thermal conductivity decreases with the increase in the interior braiding angle. For periodic microstructure unit cell the periodic boundary conditions should be imposed rather than adiabatic one.

**Acknowledgment:**

This study is supported by the Key Project of International Joint Research of NNSFC (51320105004).